\documentstyle[epsf,fleqn,aps]{revtex}
\def\Section#1{}
\def\beq{\begin{equation}}
\def\eeq{\end{equation}}
\def\bea{\begin{eqnarray}}
\def\eea{\end{eqnarray}}
\def\nn{\nonumber}

\begin{document}
\tolerance 50000
\twocolumn[\hsize\textwidth\columnwidth\hsize\csname
@twocolumnfalse\endcsname

\title{Quasi-periodic spin chains in a magnetic field}
\author{M.\ Arlego$^1$, D.C.\ Cabra$^{1,2}$ and M.D.\ Grynberg$^1$}

\address{$^1$Departamento de F\'{\i}sica, Universidad Nacional de La Plata,
C.C.   67, (1900) La Plata, Argentina.\\
$^2$Facultad de Ingenier\'{\i}a, Universidad Nacional de Lomas de Zamora,\\
Cno. de Cintura y Juan XXIII, (1832), Lomas de Zamora, Argentina.}

\maketitle

\begin{abstract}

We study the interplay between a (quasi) periodic coupling array
and an external magnetic field in a spin-$\frac{1}{2}$ $XXZ$
chain. A new class of magnetization plateaux are obtained by means
of Abelian bosonization methods which give rise to a sufficient
quantization condition. The investigation of magnetic phase
diagrams via exact diagonalization of finite clusters finds a
complete agreement with the continuum treatment in a variety of
situations. \vskip 0.5cm

PACS numbers: 71.10.Fd, \,71.10.Pm,\, 75.60.Ej.
\end{abstract}


\vskip -0.2cm \vskip2pc]

The magnetic properties of quasi-crystals have become a fundamental
issue of study since their discovery in 1984 \cite{SBGC}.
A variety of theoretical efforts, ranging from renormalization
group (RG) analysis of Ising models in Penrose lattices \cite{GLO}
to exact solutions of both Ising and $XY$ Fibonacci spin
chains \cite{ALM,LN,KST}
have revealed fairly intricate magnetic orderings associated
to the quasi-periodicity of these structures.
The non-metallic spin exchange mechanism
implicit in those studies has been evidenced in
recently synthesized rare earth ($\cal R$) ZnMg-$\cal R$
quasi-crystals (see {\it e.g.} \cite{Setal})
whose $\cal R$ elements have well localized $4f$
magnetic moments.

Bolstered by these latter findings and as a further step within
the line of the local moment descriptions referred to above, here
we consider the ordering of quasi-periodic spin-$\frac{1}{2}\,$
$XXZ$ chains in a magnetic field to elucidate the
quantization conditions of massive spin excitations or
magnetization plateaux. In periodic systems, this issue
has received systematic attention in the last few years from both
experimental and theoretical points of view (see {\it e.g.} \cite{review}).
In this letter, we are specifically interested in studying
the antiferromagnetic system
\begin{eqnarray}
\nonumber
H_{qp} = &J& \sum_n \left(1+\epsilon_n \right)\,
\left(\,S^x_n \, S^x_{n+1} \,+ \, S^y_n \, S^y_{n+1}\right.\\
 &+& \left. \Delta \,S^z_n \,S^z_{n+1} \,\right )\,-\, h\, \sum_n S^z_n\,,
\label{qp}
\end{eqnarray}
where $S^x,S^y,S^z\,$ denote the spin-$\frac{1}{2}$ matrices involved
in the standard  $XXZ$ Hamiltonian ($\epsilon_n = 0\,$)
in a magnetic field $h$ applied along the anisotropy direction
$(\,\vert \Delta \vert \le 1\,)$.
Here, the coupling modulation is introduced via the $\epsilon_n$
parameters defined as $\epsilon_n = \sum_{\nu} \delta_{\nu} \,
\cos \left( 2 \pi\, \omega_{\nu} n \,\right)\,$, so quasi-periodicity
arises upon choosing an irrational subset of
frequencies $\omega_{\nu}$  with amplitudes $\delta_{\nu}\,$.

The interest of (\ref{qp}) stems partly from the widespread
applications of $1d$ Hamiltonians in the description of
artificially grown quasi-periodic heterostructures \cite{Merlin},
quantum dot crystals \cite{Kouwen} and magnetic multilayers \cite{BACA}.
Also, recent investigations of quasi-periodicity involving either
the couplings \cite{VMG} or the magnetic field \cite{Hida2}, have been
addressed using Abelian bosonization  along with RG and numerical
techniques. Here we focus on the combined effect of a quasi-periodic
exchange modulation under a uniform magnetic field.

Of particular importance are the rational frequencies of
(\ref{qp}), not only as a way to approach the quasi-periodic
limit, but also because they allow for a thorough numerical
verification of a novel situation (see \cite{CHN,Wetal,DRZ} for
related work). As we shall see, although the allowed fractional
plateaux predicted in the present case fall into the
classification provided by the generalized Lieb-Schultz-Mattis
theorem \cite{AOY}, a bosonization approach to (\ref{qp}) yields
an alternative scenario not envisaged in previous studies
\cite{TotsukaOne,CHP}. This will be reflected in the appearance of
magnetization plateaux associated to each of the frequencies
present in (\ref{qp}). To strengthen the potential interest of our
results, we show how a simple two frequency model exhibits a
magnetization curve with two wide plateaux at 1/4 and 3/4 of
saturation, a situation which is highly reminiscent of that
observed in magnetization experiments on NH$_4$CuCl$_3$
\cite{Tanaka}.

Following the standard bosonization procedure (see {\it e.g.} \cite{book}),
the continuum limit of the $XXZ$ Hamiltonian in the presence of an
external magnetic field $h$ is given by the Tomonaga-Luttinger
Hamiltonian
\begin{equation}
H = {\frac{1 }{2}} \int dx \left( v K (\partial_x \tilde{\phi})^2 +
{\frac{v}{K}} (\partial_x \phi)^2 \right) \, .  \label{Hbos}
\end{equation}
The bosonic field $\phi$ and its dual $\tilde{\phi}$ are given by
the sum and difference of the light-cone components, respectively.
The constant $K=K(\langle M \rangle , \Delta)$ governs the
conformal dimensions of the bosonic vertex operators and can be
obtained exactly from the Bethe Ansatz solution of the $XXZ$ chain
(see {\it e.g.} \cite{CHP} for a detailed summary). One has $K=1$ for
the $SU(2)$ symmetric case ($\Delta = 1$) and it is related to the
radius $R$ of \cite{CHP} by $K^{-1} = 2 \pi R^2$.
In terms of these fields, the spin operators read
\begin{equation}
\hspace{-0.6cm}
S_x^z ={\frac 1{\sqrt{2\pi }}}\partial _x\phi +a :\cos (2k_Fx+%
\sqrt{2\pi }\phi ):+\frac{\langle M\rangle }2\,,  \label{EQsz}
\end{equation}
\begin{equation}
\hspace{-0.6cm}
S_x^{\pm } =(-1)^x:{\rm e}^{\pm i\sqrt{2\pi }\tilde \phi }\left( b\cos
(2k_Fx+\sqrt{2\pi }\phi )+c\right) :\,,  \label{EQsp}
\end{equation}
where the colons denote normal ordering with respect to the
ground state with magnetization $\langle M\rangle $. The Fermi
momentum $k_F$ is related to the magnetization of the chain as
$k_F=(1-\langle M\rangle )\pi /2$. Either the $XXZ$ anisotropy or an
external magnetic field modify the scaling dimensions of the
physical fields through $K$ and the commensurability properties of
the spin operators, as can be seen from (\ref{EQsz}),
(\ref{EQsp}). The non-universal constants $a$, $b$ and $c$ can be
in general computed numerically (see {\it e.g.} \cite{HF}, for the case
of zero magnetic field) and in particular the constant $b$ has
been obtained exactly in \cite{LZ}.

For the sake of clarity, let us first consider the effect
introduced by a {\it single} frequency term in Eq.\,(\ref{qp}),
i.e. $\epsilon_n =  \delta\,\cos \left( 2 \pi\, \omega \, n
\right)\,$. Thus, using Eqs.\ (\ref{EQsz}) and (\ref{EQsp}) it
follows that the relevant part of the interaction term in $H_{qp}
= H + H_{int}$ reads
\begin{eqnarray}
H_{int} =\sum_x \cos \left( 2\pi \omega x\right) \left[
\lambda _1(\partial _x\tilde \phi )^2+\lambda _2(\partial_x\phi )^2
\right.\nn\\
\hspace{-2cm}
\left.
+\,\lambda _3\cos (2k_Fx+\sqrt{2\pi }\phi)\,+\,
\lambda _4\sin (2k_Fx+\sqrt{2\pi }\phi )\right]
\label{EQHint}
\end{eqnarray}
where $\lambda_i \propto \delta\,, ~ i=1,\,\cdots\,,4\,$.

As in previous analysis \cite{AOY,TotsukaOne,CHP} one can readily
obtain the necessary quantization condition for the appearance of
a plateau by looking at the commensurability of the relevant
operators. In the present case we need only to consider the vertex
operators $\exp \pm i \sqrt{2\pi} \phi$ of scaling dimension
$d=1/(4\pi R^2)$. Therefore, we obtain that $\langle M \rangle$
should satisfy
\beq
\langle M \rangle = \pm (2 \omega -1)\ ,
\label{quantcond}
\eeq
in order for a plateau to be present. The
fulfillment of this condition opens a spin gap excitation since
the operator in question is relevant at least for $0 <\Delta < 1$.
In the region $-1 < \Delta < 0$ a critical curve appears which can
be determined from the Bethe-Ansatz solution for $R(\langle M
\rangle , \Delta)$ \cite{CHP}. Furthermore, the gap width can be
easily computed to scale as $\delta^{1/(2-d)}$.
Notice that the perturbations $\lambda_{1,2}$ do not play a
role here since they are incommensurate whenever (\ref{quantcond}) holds.

In fact, these expectations turn out to apply well above the weak
coupling regime discussed so far. In Fig.\ 1 we show the
magnetization phase diagram resulting from exact diagonalization
of fairly large $XX$ chains ($10^5$ spins)  by setting $\omega =
13/21\,$ through a wide range of couplings. Notice that for
rational frequencies, $\omega \in {\cal Q}\,$ and $\delta \to
\delta_c\,$, for some appropriate value of $\delta_c$ which
depends on $\omega$, the chain breaks up into a periodic
collection of {\it finite} segments which naturally yield
additional plateaux of rational values. Being the unit cell
composed of $21$ spins one would naively expect plateaux to appear
for values of $M= (2n+1)/21$, $n=0,\cdots,10$, but as can be seen
from Fig.\ 1 this is not the case due to the non-trivial structure
of the unit cell and only some of these values are indeed
observed. Interestingly, Eq.\,(\ref{quantcond}) yet remains robust
all the way through the decoupling point and the corresponding
plateau corresponds to the most prominent.

These observations were also corroborated on $XXZ$ chains. However,
owing to the large spaces  involved in their diagonalization
now we content ourselves with moderate lengths $L$ (up to 24 spins).
After resorting to the Lanczos method \cite{Lanczos}
in each of the magnetization subspaces with $S^z \in
\{0,\,1,\,\cdots\,,L/2\}\,$, we built up the magnetization
contours shown in Fig.\ 2 using $\omega = 5/8\,$ for $L=8,\,16,\,24\,$.
As expected, the massless $XXZ$ excitations around
$\langle M \rangle = 1/4\,$ render size
effects rather noticeable within small coupling regions.
Nevertheless, already for $\delta > 0.2\,$ they clearly become less
pronounced, thus lending our results further support
to the bosonization picture.

A word of caution should be added here, namely, the importance of
periodic boundary conditions (PBC) in testing the analytic
approach via exact diagonalization of small systems. Already at
the level of a simple dimerized chain ($\omega = 1/2$), PBC become
crucial. In fact, the numerical analysis of this latter situation
using {\it open} boundary conditions shows that the well known
$\langle M \rangle =0$ plateau expected in the large length
$L$-limit, actually emerges at $\langle M \rangle =2/L$. Fig.\ 3
illustrates this observation for $L=$ 24, 20 and 16 and should
emphasize the essential role of PBC in all our subsequent
numerical checks.

By construction, the bosonization approach can be straightforwardly
extended to the case in which more than one frequency is present in
$\epsilon_n$. It turns out that whenever condition
(\ref{quantcond}) is satisfied {\it for each frequency}, a
magnetization plateau shows up. This extension is
simple since each operator (corresponding to each frequency) is
commensurate separately, and hence the different perturbations can
be treated separately. Of course the situation changes in the
case of a dense
multi-frequency spectrum (such as in the Fibonacci potential that we
discuss below) and a more complete analysis has to be carried
out \cite{VMG}.

In order to test the reliability
of these predictions, we have analyzed numerically both $XX$ and
$XXZ$ chains using double-frequency couplings. In Fig. 4 we
display the magnetization curves obtained for $\omega_1 = 5/8,\,
\omega_2 = 7/8\,$ with amplitudes $\delta_1 = 0.2\,$ and
$\delta_2 = 0.3\,$ respectively. As we mentioned before, the
rather robust plateaux emerging at $\langle M \rangle =\,$ 1/4 and
3/4 not only confirm the correctness of our extended bosonization
prediction, but also pave the way to an alternative description of
the massive spin excitations observed in NH$_4$CuCl$_3$; a
quasi-one-dimensional $S=\frac{1}{2}$ compound which is attracting
both theoretical and experimental attention and whose
magnetization behavior yet remains unexplained
\cite{Tanaka,Kolezhuk,CaGyThree}. Of course, a more realistic description
of this material must start from the microscopic structure
observed by X-ray spectroscopy, which points to a two-leg zigzag
ladder. Still, it is quite encouraging to obtain a magnetization
curve qualitatively similar with the simple two-frequency model we
considered here.

In studying irrational frequencies or other quasi-periodic
modulations, it is natural to check our analysis with
the prototype Fibonacci sequence, a coupling array
$J_{A}=J(1+\delta)$, $J_{B}=J(1-\delta)$ generated
by iterating the substitution rules $B \rightarrow A$ and $A
\rightarrow AB$ \cite{ALM,LN,KST,VMG}. Here we discuss the general
$XXZ$ situation (a related model has been
studied in \cite{T}\,), and compare our results with the
already well known magnetization curve of the $XX$ case \cite{LN,KST}.
Before continuing with the bosonization
approach, we pause to discuss the {\it strong} coupling regimes of
this system ($\delta \to \pm 1$) in the context of a simple
decimation procedure (see {\it e.g.} \cite{CdMGPP}). To evaluate
the magnetization of the widest plateaux, there are two different
cases to consider, according to $\delta \simeq -1,\,$ i.e. $J_B
\gg J_A,$  and the opposite situation for $\delta \simeq 1$.

Starting from saturation, in the first case the magnetic field
is lowered  until it reaches the value $h_c
\simeq J_B$ at which the type-$B$ bonds experience a
transition from the state of maximum polarization to the singlet
state. The magnetization at this plateau is then obtained by
decimating the $B$ bonds. This simply yields $\langle M \rangle = 1 - 2
N_B/(N_B+N_A)$, where $N_{A,B}$ denotes the number of bonds of type
$A$ and $B$ respectively.  For a
large iteration number of the rules referred to above ($L \to \infty\,$),
$N_A/N_B$ approaches the golden mean $\gamma = (1+\sqrt{5})/2$ and therefore
we find $M_1 = (\gamma-1)/(\gamma+1) \simeq 0.236068$.

In the second case, $J_A \gg J_B$, we have to distinguish two
different unit cells since type-$A$ bonds can appear either in
pairs (forming trimers) or isolated (forming dimers). It can be readily
checked that when lowering the magnetic field from saturation the
first spins to be decimated correspond to those forming trimers.
We then find two plateaux at $M_2=1-2/\gamma^3 \simeq 0.527864$
(after decimating trimers) and, alike the case $\delta \simeq
1\,$, at $M_1$ (after decimating the remaining dimers).
Since the decimation procedure applies for
generic $XXZ$ chains \cite{CdMGPP}, we conclude that the emergence
of these strong coupling plateaux is a generic feature, at
least with an antiferromagnetic anisotropy parameter $0<\Delta <1$.

For intermediate regimes $0 < \vert \delta \vert < 1\,$ the
magnetization curve has a much richer structure which can be
easily understood from our bosonization analysis in a
multi-frequency case. Evidently, the self similar hierarchy of
frequencies resulting from the Fourier transform of the Fibonacci
exchanges (see {\it e.g.} \cite{VMG}) along with the quantization
condition studied so far, enables to reconstruct the whole spin
gap structure of the Fibonacci chain, at least for $\vert \delta
\vert \ll 1\,$ and as long as the operator responsible for the
plateaux is relevant. Actually, it turns out that one can obtain a
fair approximation to the latter showing the most important
plateaux by keeping just a few number of main spectrum
frequencies, even beyond weak coupling regimes. Interestingly,
when the constraint (\ref{quantcond}) is applied to the dominant
Fibonacci frequencies $\omega_1 = 1/\gamma$ and $\omega_2 =
2\,(1-1/\gamma),\,$ it yields precisely the $M_1$ and $M_2$
plateaux arising from the strong coupling decimation. These
results are shown in Fig. 5 where both single and double-frequency
approximations are displayed. Moreover, using the Fourier spectrum
to set $\delta_2/\delta_1 \simeq 0.36237,\,$ the Fibonacci gap
widths exhibit a remarkable agreement with the scaling exponents
$1/(2-d)$ referred to above. This can be observed in the inset
over more than three decades in $\delta$. In turn, this supports
the observation that {\it all} gap widths of the $XX$ Fibonacci
chain scale simultaneously with $\delta$, as for $\Delta = 0\,$
the compactification radii comprehended in the scaling exponents
are independent of the magnetization [$R \equiv 1/(2\sqrt\pi)\,$]
(see {\it e.g.} \cite{CHP}).

Finally, we point out that for the Fibonacci chain (which has a
dense Fourier spectrum), it was shown by solving the one-loop RG
equations at zero magnetic field \cite{VMG} (see also
\cite{Hida2}) that the critical value of $R$ at which the
Kosterlitz-Thouless transition occurs is moved to $R_c =
1/\sqrt{4\pi}$, in contrast to $R_c = 1/\sqrt{8\pi}$ for the
single, or more generally non-dense frequency case. Hence, for
$0<\Delta <1$ and arbitrary magnetization, the operator
responsible for spin gap openings at each of the Fibonacci
frequencies is relevant. Therefore, the magnetization curve of the
$XXZ$ case with antiferromagnetic anisotropy is of the same form
as that of the $XX$ situation (see Fig.\ 5), though with different
plateaux widths.

To summarize, we have studied the interplay between
quasi-periodic exchanges and uniform magnetic fields
in strongly correlated antiferromagnetic chains using
both bosonization and numerical techniques.
The former were tested and complemented by the latter
in a variety of non-perturbative scenarios.
Our calculations suggest the possibility to observe
rather stable magnetization plateaux [Eq.\,(\ref{quantcond})\,]
on artificially grown arrays of quantum dots \cite{Kouwen}
according to the (controlled) spatial distribution of their
exchange integrals. We trust this work will convey an interesting
motivation for further experimental studies in these
material technologies.

It is a pleasure to acknowledge useful discussions with A. Honecker
and P. Pujol. The research of D.C.C and M.D.G. is partially supported
by CONICET and Fundaci\'on Antorchas, Argentina
(grant No.\ A-13622/1-106).

\newpage
\begin{figure}
\hbox{%
\vspace{-3.5cm}}
\hbox{%
\hspace{-0.9cm}
\epsfxsize=3.8in
\epsffile{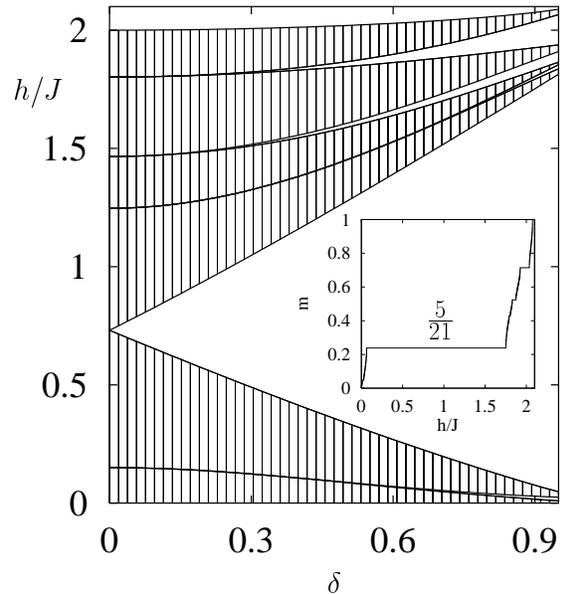}}
\vspace{-2.6cm}
\caption{Critical fields (in bold lines) of the $XX$ chain with
$\omega = 13/21$ and $10^5$ spins. At $\delta = 0.9\,$
these fields conform the standard magnetization curve displayed
in the inset. Vertical lines denote regions of massless spin excitations
where the magnetization increases continuously with $h$.
Empty zones in ascending order represent plateaux appearing at
$\langle M \rangle = \,$ 1/21,  5/21 [\,central region expected
by Eq.\,(6)\,], 9/21, 11/21  and 15/21 before saturation (top zone).}

\end{figure}

\begin{figure}
\hbox{%
\vspace{-2.5cm}}
\hbox{%
\hspace{-0.3cm}
\epsfxsize=3.8in
\epsffile{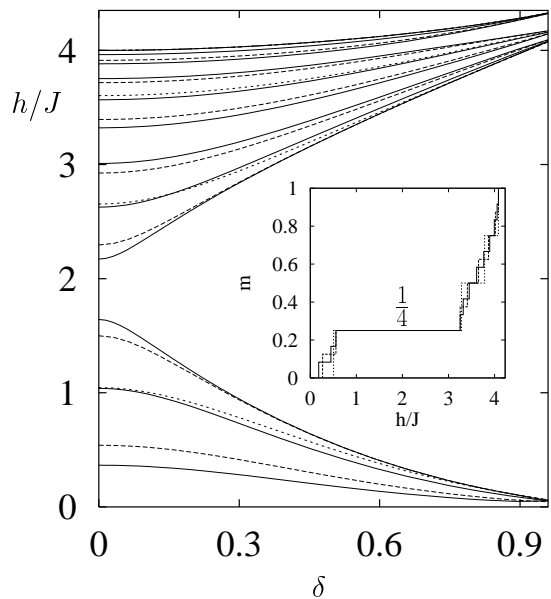}}
\vspace{-2.5cm}
\caption{Magnetization contours of the isotropic ($\Delta =1$)
$XXZ$ chain for $\omega = 5/8\,$. Solid, dashed and
dotted lines stand respectively for the critical fields
of $L=24,\, 16\,$ and 8. The middle empty region
corresponds to the $\langle M \rangle = 1/4\,$
plateau expected by Eq.\,(6). The inset shows
one of the typical magnetization curves upon setting
$\delta = 0.5\,$.}
\end{figure}

\begin{figure}
\hbox{%
\vspace{-1cm}}
\hbox{%
\hspace{-1.4cm}
\epsfxsize=3.4in
\epsffile{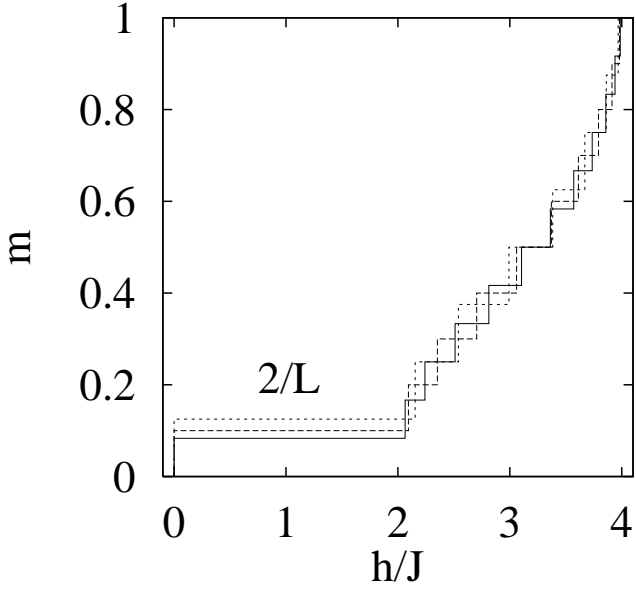}}
\vspace{0.1cm}
\caption{Magnetization curves of a dimerized Heisenberg chain
($\omega = 1/2\,$ with $\delta = 0.4$),  using {\it open} boundary
conditions displaying a plateau at $\langle M \rangle = 2/L\,$.
Solid, dashed  and dotted lines denote respectively the results of
$L =$  24, 20 and 16.}
\end{figure}


\begin{figure}
\hbox{%
\vspace{-5.3cm}}
\hbox{%
\hspace{-1.8cm}
\epsfxsize=4.7in
\epsffile{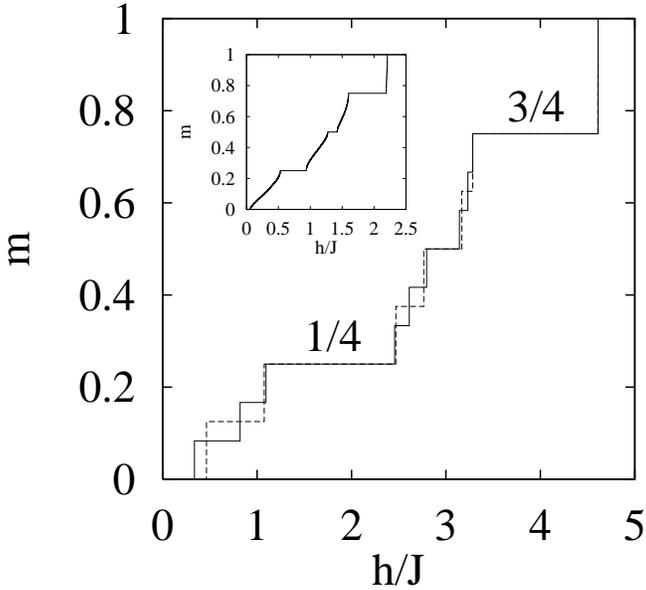}}
\vspace{-2.cm}
\caption{Double frequency magnetization curve of the isotropic
($\Delta =1$) $XXZ$ chain  for $\omega_1 = 5/8,\, \omega_2 = 7/8\,$
with amplitudes $\delta_1 = 0.2\,$ and  $\delta_2 = 0.3\,$.
Solid and dashed lines denote respectively
the  magnetizations of $L = 24$ and 16. The inset shows
the magnetization curve of the corresponding $XX$ chain with
$10^5\,$ spins.}
\end{figure}

\begin{figure}
\hbox{%
\vspace{-6.8cm}}
\hbox{%
\hspace{-1.5cm} \epsfxsize=4.7in \epsffile{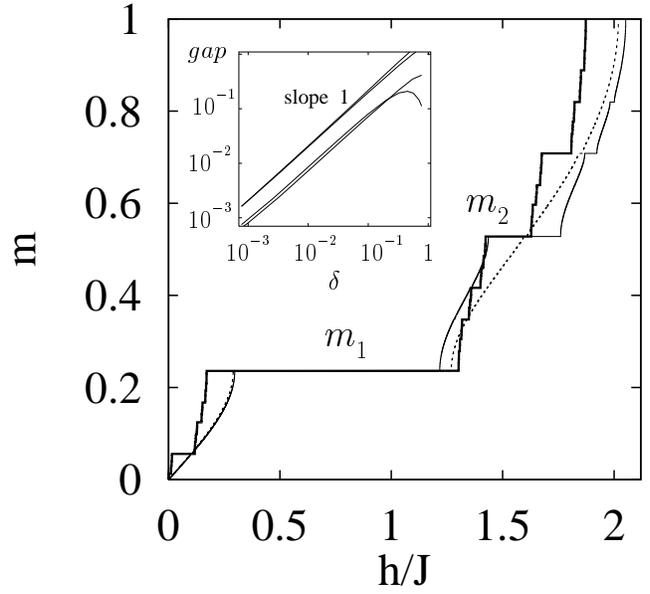}}
\vspace{-2.cm} \caption{Fibonacci magnetization for $\delta = 0.5$
and $F(25) = 75025\,$ $XX$ spins (bold line). Here, $F(25)$
denotes the $25^{th}$ Fibonacci number defined as
$F(n+1)=F(n)+F(n-1)$ with $F(1)=F(2)=1$. Dotted and solid lines
stand respectively for the single and double frequency
approximants to the $M_1$ and $M_2$ plateaux referred to in the
text. The slopes compared in the inset show the gap width of $M_2$
and $M_1$, in ascending order. Bold lines denote the Fibonacci
gaps whereas solid lines represent the double approximant widths.}
\end{figure}
\end{document}